# The Five-Minute Rule Ten Years Later, and Other Computer Storage Rules of Thumb

Jim Gray
Goetz Graefe

September 1997

Microsoft Research
Microsoft Corporation
One Microsoft Way
Redmond, WA 98052

# The Five-Minute Rule Ten Years Later, and Other Computer Storage Rules of Thumb

Jim Gray, Goetz Graefe
Microsoft Research, 301 Howard St. #830, SF, CA 94105
{Gray, GoetzG}@Microsoft.com

**Abstract:**

Simple economic and performance arguments suggest appropriate lifetimes for main memory pages and suggest optimal page sizes. The fundamental tradeoffs are the prices and bandwidths of RAMs and disks. The analysis indicates that with today's technology, five minutes is a good lifetime for randomly accessed pages, one minute is a good lifetime for two-pass sequentially accessed pages, and 16 KB is a good size for index pages. These rules-of-thumb change in predictable ways as technology ratios change. They also motivate the importance of the new *Kaps*, *Maps*, *Scans*, and *$/Kaps*, *$/Maps*, *$/TBscan* metrics.

## 1. The Five-Minute Rule Ten Years Later

All aspects of storage performance are improving, but different aspects are improving at different rates. The charts in Figure 1 roughly characterize the performance improvements of disk systems over time. The caption describes each chart.

In 1986, randomly accessed pages obeyed the *five-minute rule* [1]: pages referenced every five minutes should have been kept in memory rather than reading them from disk each time. Actually, the break-even point was 100 seconds but the rule anticipated that future technology ratios would move the break-even point to five minutes.

The five-minute rule is based on the tradeoff between the cost of RAM and the cost of disk accesses. The tradeoff is that caching pages in the extra memory can save disk IOs. The break-even point is met when the rent on the extra memory for cache ($/page/sec) exactly matches the savings in disk accesses per second ($/disk_access/sec). The break even time is computed as:

*BreakEvenReferenceInterval (seconds) =*

$$\frac{PagesPerMBofRAM}{AccessPerSecondPerDisk} \times \frac{PricePerDiskDrive}{PricePerMBofDRAM} \qquad (1)$$

The disk price includes the cost of the cabinets and controllers (typically 30% extra.) The equations in [1] were more complex because they did not realize that you could factor out the depreciation period.

The price and performance from a recent DELL TPC-C benchmark [2] gives the following parameters for Equation 1:

*PagesPerMBofRAM* = 128 pages/MB (8KB pages)
*AccessesPerSecondPerDisk* = 64 access/sec/disk
*PricePerDiskDrive* = 2000 $/disk (9GB + controller)
*PricePerMBofDRAM* = 15 $/MB_DRAM

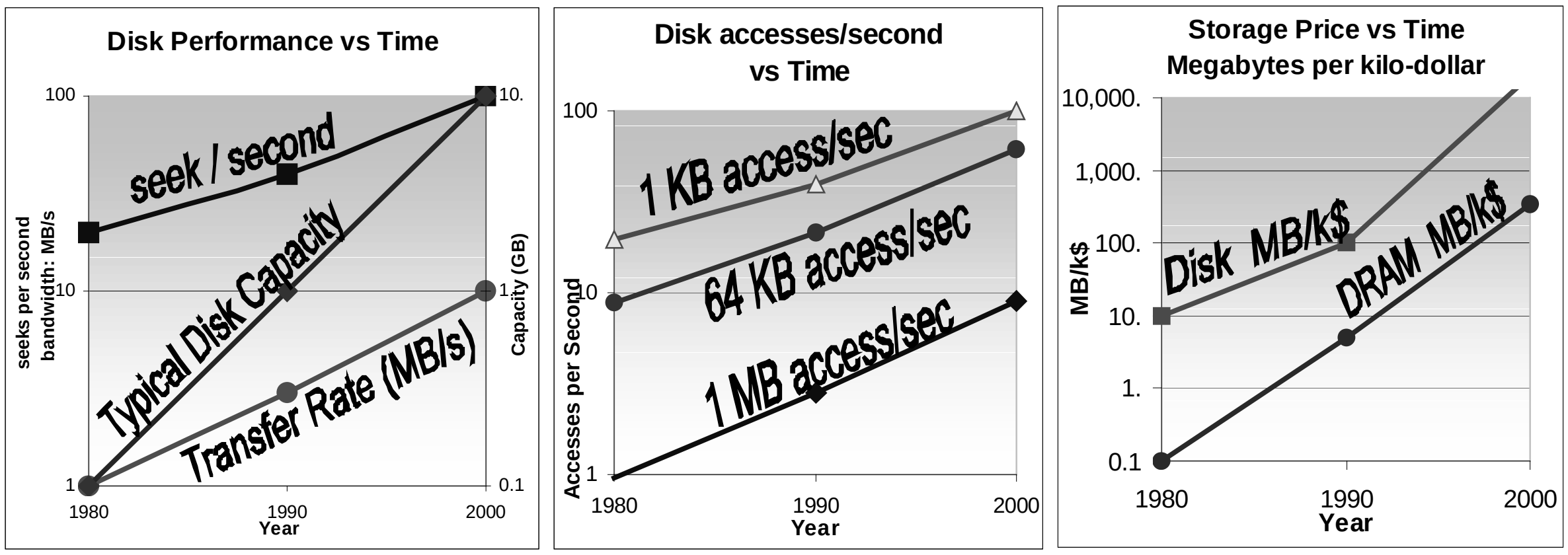

**Figure 1:** Performance of magnetic storage disks over time. The first two graphs show that accesses and access times improved 10x or 100x while capacity grew 100x. The third graph shows that prices improved 1,000x in the same time. We have compensated for the changing ratios among accesses, capacity, and cost by using larger RAM buffers and larger pages. That is one theme of this paper.

Evaluating Equation 1 with these values gives a reference interval of 266 seconds -- about five minutes[1]. So, even in 1997, data referenced every five minutes should be kept in main memory.

Prices for the same equipment vary enormously, but all the categories we have examined follow something like a five-minute rule. Server hardware prices are often three times higher than "street prices" for the same components. DEC Polaris RAM is half the price of DELL. Recent TPC-C Compaq reports have 3x higher RAM prices (47$/MB) and 1.5x higher disk prices (3129$/drive) giving a two-minute rule. The March 1997 SUN+Oracle TPC-C benchmark [3] had prices even better than DELL (13$/MB of RAM and 1690$ per 4GB disk and controllers). These systems all are near the five-minute rule. Mainframes are at 130$/MB for RAM, 10K$/MIPS, and 12k$/disk. Thus, mainframes follow a three-minute rule.

One can think of the first ratio of Equation 1 (*PagesPerMBofRAM/AccessesPerSecondPerDisk*) as a *technology ratio*. The second ratio of Equation 1 (*PriceofDiskDrive/PriceOfMBofRAM*) is an *economic ratio*. Looking at the trend lines in Figure 1, the technology ratio is shifting. Page size has increased with accesses/second so the technology ratio has decreased ten fold (from 512/30 = 17 to 128/64 = 2). Disk drive prices dropped 10x and RAM prices dropped 200x, so that the economic ratio has increased ten fold (20k$/2k$=10 to 2k$/15$=133). The consequent *reference interval* of equation (1) went from 170 seconds (17x10) to 266 seconds (2x133).

**These calculations indicate that the *reference interval* of Equation (1) is almost unchanged, despite these 10x, 100x, and 1,000x changes. It is still in the 1-minute to 10-minute range. The 5-minute rule still applies to randomly accessed pages.**

The original paper [1] also described the 10-byte rule for trading CPU instructions off against DRAM. At the time one instruction cost the same as 10 bytes. Today, PCs follow a 1-byte rule, mini-computers follow a 10 byte rule, while mainframes follow a kilobyte rule because the processors are so overpriced.

[1] The current 2 KB page-size of Microsoft SQL Server 6.5 gives a reference interval of 20 minutes. MS SQL is moving to an 8 KB page size in the 1998 release.

### 1.2. Sequential Data Access: the One-Minute Sequential Rule

The discussion so far has focused on random access to small (8KB) pages. Sequential access to large pages has different behavior. Modern disks can transfer data at 10 MBps if accessed sequentially (Figure 1a). That is a peak value, the analysis here uses a more realistic 5 MB/s as a disk sequential data rate. Disk bandwidth drops 10x (to 0.5 MBps) if the application fetches random 8KB pages from disk. So, it should not be surprising that sequential IO operations like sort, cube, and join, have different RAM/disk tradeoffs. As shown below, they follow a one-minute-sequential rule.

If a sequential operation reads data and never references it, then there is no need to cache the data in RAM. In such one-pass algorithms, the system needs only enough buffer memory to allow data to stream from disk to main memory. Typically, two or three one-track buffers (~100 KB) are adequate. For one-pass sequential operations, less than a megabyte of RAM per disk is needed to buffer disk operations and allow the device to stream data to the application.

Many sequential operations read a large data-set and then revisit parts of the data. Database join, cube, rollup, and sort operators all behave in this way. Consider the disk access behavior of Sort in particular. Sort uses sequential data access and large disk transfers to optimize disk utilization and bandwidth. Sort ingests the input file, reorganizes the records in sorted order, and then sequentially writes the output file. If the sort cannot fit the file in main memory, it produces sorted runs in a first pass and then merges these runs into a sorted file in the second pass. Hash-join has a similar one-pass two-pass behavior.

The memory demand of a two pass sort is approximately given in equation 2:

$$MemoryForTwoPassSort \approx 6 \times Buffer_Size + \sqrt{3 \times Buffer_Size \times File_Size} .... (2)$$

Equation 2 is derived as follows. The first sort pass produces about *File_Size/Memory_Size* runs while the second pass can merge *Memory_Size/Buffer_Size* runs. Equating these two values and solving for memory size gives the square root term. The constants (3 and 6) depend on the particular sort algorithm. Equation 2 is graphed in Figure 2 for file sizes from megabytes to exabytes.

Sort shows a clear tradeoff of memory and disk IO. A one-pass sort uses half the disk IO but much more memory. When is it appropriate to use a one-pass sort? This is just an application of Equation 1 to compute the break-even reference interval. Use the DEC TPC-C prices [2] and components in the previous section. If sort uses to 64KB transfers then there are 16 pages/MB and it gets 80 accesses per second (about 5 MB/s).

*PagesPerMBofRAM* = 16 pages/MB
*AccessesPerSecondPerDisk* = 80 access/sec/disk

Using these parameters, Equation 1 yields a break-even reference interval of 26 seconds (= (16/80) x (2,000/15)). Actually, sort would have to write and then read the pages, so that doubles the IO cost and moves the balance point to 52 seconds. Anticipating higher bandwidths and less expensive RAM, we predict that this value will slowly grow over time.

Consequently, we recommend the ***one-minute-sequential rule****: hash joins, sorts, cubes, and other sequential operations should use main memory to cache data if the algorithm will revisit the data within a minute.*

For example, a one-pass sort is known to run at about 5 GB/minute [4]. Such sorts use many disks and lots of RAM but they use only half the IO bandwidth of a two-pass sort (they pass over the data only once). Applying the one-minute-sequential rule, below 5 GB a one-pass sort is warranted. Beyond that size, a two-pass sort is warranted. With 5GB of RAM a two-pass sort can sort 100 terabytes. This covers ALL current sorting needs.

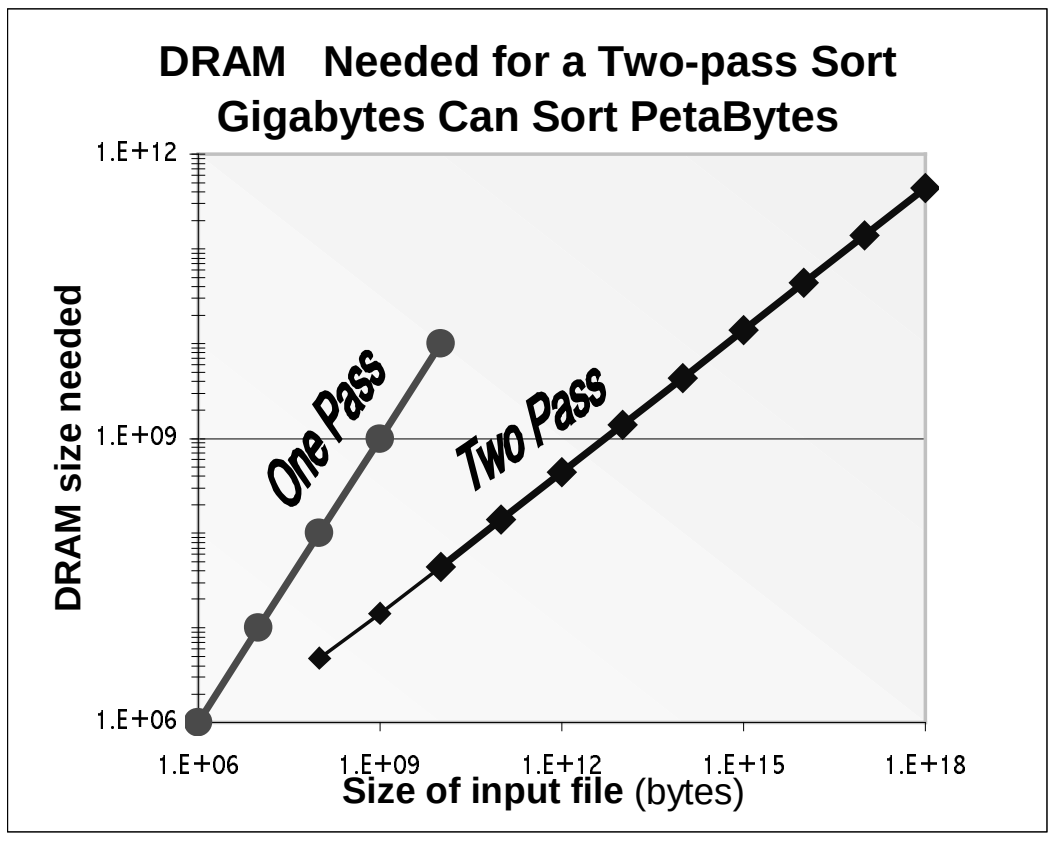

**Figure 2:** A two-pass sort can process 100 terabyte files with a 5 GB DRAM buffer. The two pass sort balances the run length against the number of runs to merge in the second pass. If it generates a thousand runs of 100 MB each, it can merge them using 100 MB of merge buffers in phase 2. This is a 100 GB sort. With current technology, use a 1-pass sort up to 5GB files. For larger files, do a 2-pass

Similar comments apply to other sequential operations (group by, rollup, cube, hash join, index build, etc…). **In general, sequential operations should use high-bandwidth disk transfers and they should cache data that they will revisit the data within a minute.**

In the limit, for large transfers, sequential access cost degenerates to the cost of the bandwidth. The technology ratio of equation 1 becomes the reciprocal of the bandwidth (in megabytes):

*TechnologyRatio*
*= (PagesPerMB)/(AccessesPerSecond)*
*= (1E6/TransferSize)/*
*( DiskBandwidth/TransferSize)*
for purely sequential access
*= 1E6/DiskBandwidth.* *(3)*

This is an interesting result. It gives rise to the asymptote in Figure 3 that shows the reference interval vs. page size. With current disk technology, the reference interval asymptotically approaches 40 seconds as the page size grows.

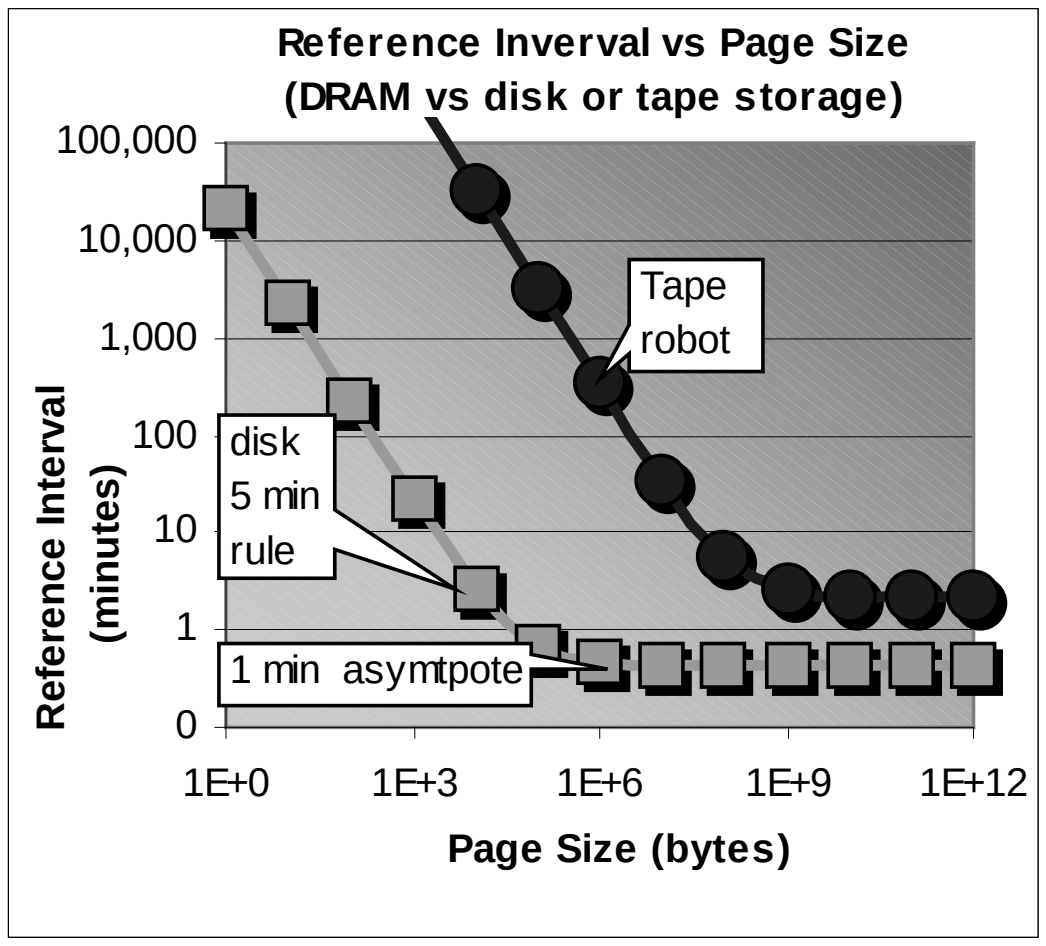

**Figure 3:** The break-even reference interval for disk vs. DRAM asymptotically approaches something like one minute for current technology. The asymptote is the product of the technology ratio (which becomes 1e6/bandwidth) and the economic ratio. A later section discuses the disk-tape tradeoff. Fundamentally, tape technology is VERY expensive to access. This encourages very large tape page sizes and very cold data on tape. The tape asymptote is approached at 10 GB (tape hardware is described in Table 4).

## 1.4. RAID and Tape

RAID 0 (striping) spreads IO among disks and so makes the transfer size smaller. Otherwise, RAID 0 does not perturb this analysis. RAID 1 (mirroring) slightly decreases the cost of reads and nearly dou-bles the cost of writes. RAID 5 increases the cost of writes by up to a factor of 4. In addition RAID5 controllers usually carry a price premium. All these factors tend to increase the economic ratio (making disks more expensive, and raise the technology ratio (lower accesses per second). Overall they tend to increase the random access reference interval by a factor of 2x to 5x.

Tape technology has moved quickly to improve capacity. Today the Quantum DLTstor™ is typical of high performance robots. Table 4 presents the performance of this device.

**Table 4:** Tape robot price and performance characteristics (source Quantum DLTstor™).

| | |
|---|---|
| Quantum DLT Tape Robot | 9,000$ price |
| Tape capacity | 35 GB |
| Number of tapes | 14 |
| Robot Capacity | 490 GB |
| Mount time (rewind, un-mount, put, pick, mount, position) | 30 seconds |
| Transfer rate | 5 MBps |

Accessing a random data record on a tape requires mounting it, moving to the right spot and then reading the tape. If the next access is on another tape and so one must rewind the current tape, put it away, pick the next one, scan to the correct position, and then read. This can take several minutes, but the specifications above charitably assumed it takes 30 seconds on average.

When should you store data on tape rather than in RAM? Using Equation 1, the break-even reference interval for a 8KB tape block is about two months (keep the page in RAM rather than tape if you will revisit the page within 2 months).

Another alternative is keeping the data on disk. What is the tradeoff of keeping data on disk rather than on tape? The tradeoff is that tape-space rent is 10x less expensive but tape accesses are much more expensive (100,000x more for small accesses and 5x more for large (1GB) accesses). The reference interval balances the lower tape rent against the higher access cost. The resulting curve is plotted in Figure 3.

## 1.5. Checkpoint Strategies In Light of the 5-minute Rule

Buffer managers typically use an LRU or Clock2 (two round clock) algorithm to manage the buffer pool. In general, they flush (write to disk) pages when (1) there is **contention** for cache space, or (2) the page must be **checkpoint**ed because the page has been dirty for a long time. The checkpoint interval is typically five minutes. Checkpoint limits recovery to redoing the last five or ten minutes of the log.

Hot-standby and remote-disaster-recovery systems reduce the need for checkpoints because they continuously run recovery on their version of the database and can take over within seconds. In these disaster-tolerant systems, checkpoints can be very infrequent and almost all flushes are contention flushes.

To implement the *N*-minute rule for contention flushes and evictions, the buffer manager keeps a list of the names of all pages touched within the last *N* minutes. When a page is re-read from disk, if it is in the *N*-minute list, it is given an *N*-minute lifetime (it will not be evicted for *N*-minutes in the future). This simple algorithm assures that frequently accessed pages are kept in the pool, while pages that are not re-referenced are aggressively evicted.

## 1.6. Five-Minute Summary

In summary, the five-minute rule still seems to apply to randomly accessed pages, primarily because page sizes have grown from 1KB to 8KB to compensate for changing technology ratios. For large (64KB pages) and two-pass sequential access, a one-minute rule applies today.

# 2.How Large Should Index Pages Be?

The size of an internal index page determines its retrieval cost and fan-out (*EntriesPerPage)*. A B-tree indexing *N* items will have a height (in pages) of:

$$Indexheight \sim \log_2(N)/\log_2(EntriesPerPage) \text{ pages} \quad (4).$$

The *utility* of an index page measures how much closer the index page brings an associative search to the destination data record. It tells how many levels of the binary-tree fit on a page. The utility is the divisor of the Equation 4:

$$IndexPageUtility = \log_2(EntriesPerPage) \quad (5)$$

For example, if each index entry is 20 bytes, then a 2 KB index page that is 70% full will contain about 70 entries. Such a page will have a utility of 6.2, about half the utility of a 128 KB index page (see Table 6).

Reading each index page costs a logical disk access but each page brings us *IndexPageUtility* steps closer to the answer. This cost-benefit tradeoff gives rise to an optimal page size that balances the *IndexPageAccessCost* and the *IndexPageUtility* of each IO.

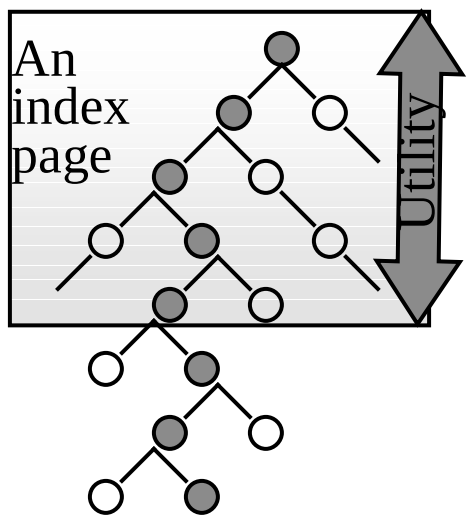

**Figure 5**: The utility of an index page is the number of levels of the binary tree that it traverses.

The utility rises as the log of the page size. The cost of the access goes up linearly with page sizeConsequently, for a particular disk latency and transfer rate, there is an optimal index page size. The tree at left shows just the search path (it is not balanced because the drawing would be too cluttered).

Reading a 2 KB page from a disk with a 10 ms average access time (seek and rotation) and 10 MB/s transfer rate uses 10.2 ms of disk device time. So the read cost is 10.2 milliseconds. More generally, the cost of accessing an index page is either the storage cost in main memory if the page is cached there, or the access cost if the page is stored on disk. If pages near the index root are cached in main memory, the cache saves a constant number of IOs on average. This constant can be ignored if one is just optimizing the IO subsystem. The index page disk access cost is

*IndexPageAccessCost = Disk Latency + PageSize / DiskTransferRate (6)*

The benefit-cost ratio of a certain page size and entry size is the ratio of the two quantities.

*IndexPageBenefit/Cost = IndexPageUtility / IndexPageAccessCost. (7)*

The right column of Table 6 shows this computation for various page sizes assuming 20-byte index entries. It indicates that 8 KB to 32 KB pages are near optimal for these parameters.

Figure 7 graphs the benefit/cost ratios for various entry sizes and page sizes for both current, and next-generation disks. The graphs indicate that, small pages have low benefit because they have low utility and high fixed disk read costs. Very large pages also have low benefit because utility grows only as the log of the page size, but transfer cost grows linearly with page size.

**Table 6:** Tabulation of index page utility and benefit/cost for 20 byte index entries assuming each page is 70% full and assuming a 10ms latency 10 MBps transfer rate.

| page size KB | entries per page Fan-out | Index Page Utility | Index Page Access Cost (ms) | Index Page Benefit/ Cost (20B) |
|---|---|---|---|---|
| 2 | 68 | 6.1 | 10.2 | 0.60 |
| 4 | 135 | 7.1 | 10.4 | 0.68 |
| 8 | 270 | 8.1 | 10.8 | 0.75 |
| 16 | 541 | 9.1 | 11.6 | 0.78 |
| 32 | 1081 | 10.1 | 13.2 | 0.76 |
| 64 | 2163 | 11.1 | 16.4 | 0.68 |
| 128 | 4325 | 12.1 | 22.8 | 0.53 |

Table 6 and Figure 7 indicate that for current devices, index page sizes in the range of 8 KB to 32 KB are preferable to smaller and larger page sizes. By the year 2005, disks are predicted to have 40 MB/s transfer rates and so 8 KB pages will probably be too small.

## 3. New Storage Metrics

These discussions point out an interesting phenomenon -- the fundamental storage metrics are changing. Traditionally, disks and tapes have been rated by capacity. As disk and tape capacity approach infinity (50 GB disks and 100 GB tapes are in beta test today), the cost/GB goes to zero and the cost/access becomes the dominant performance metric.

The traditional performance metrics are:

**GB:** storage capacity in gigabytes.

**$/GB:** device price divided by capacity.

**Latency:** time between issue of IO and start of data transmission.

**Bandwidth:** sustained transfer rate from the device.

The latter two are often combined as a single access time metric (time to read a random KB from the device).

**Kaps :** kilobyte accesses per second.

As device capacities grow, additional metrics become important. Transfers become larger. Indeed, the minimum economical tape transfer is probably a one MB object

Increasingly, applications use a dataflow style of programming and stream the data past the device. Data mining applications and archival applications are the most common example of this today. These suggest the following two new storage metrics.

**Maps:** Megabyte accesses per second.

**Scan:** how long it takes to sequentially read or write all the data in the device?

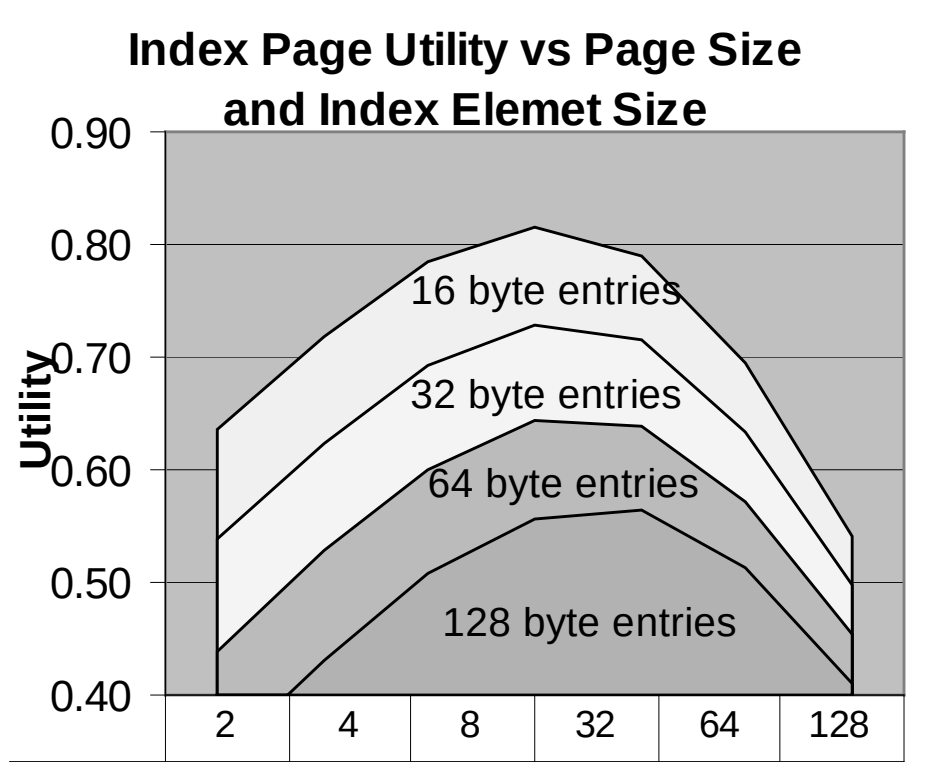

| | 2 | 4 | 8 | 32 | 64 | 128 |
|---|---|---|---|---|---|---|
| 16 B | 0.6355 | 0.7191 | 0.7843 | 0.7898 | 0.6938 | 0.5403 |
| 32 B | 0.5375 | 0.623 | 0.6919 | 0.7144 | 0.6334 | 0.497 |
| 64 B | 0.4395 | 0.527 | 0.5994 | 0.6391 | 0.573 | 0.4538 |
| 128 B | 0.3415 | 0.4309 | 0.507 | 0.5638 | 0.5126 | 0.4105 |

Page Size (KB)

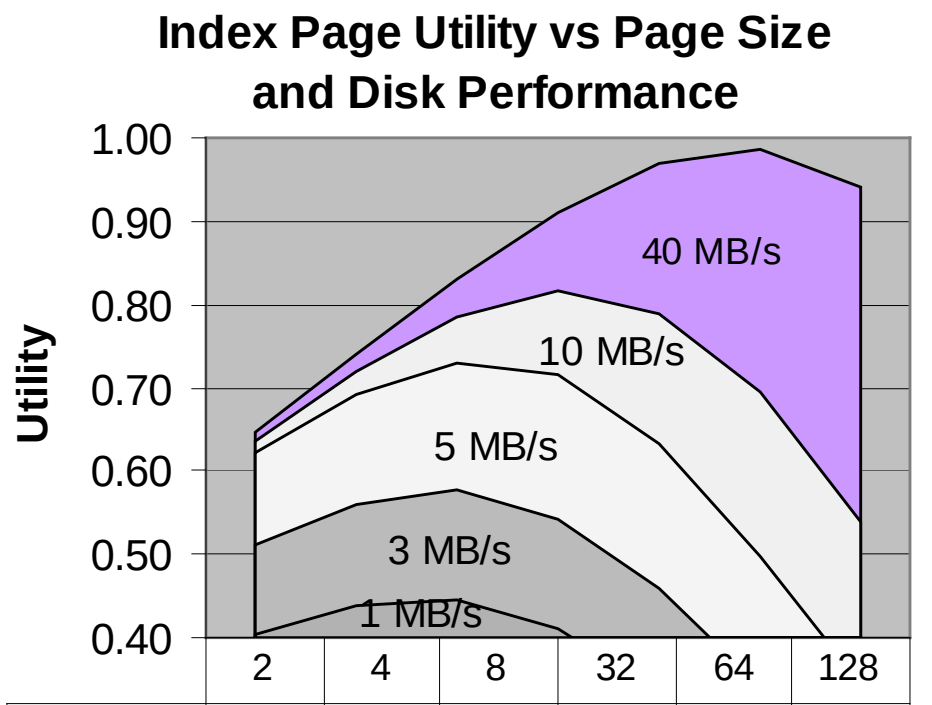

| | 2 | 4 | 8 | 32 | 64 | 128 |
|---|---|---|---|---|---|---|
| 40 MB/s | 0.645 | 0.741 | 0.832 | 0.969 | 0.987 | 0.94 |
| 10 MB/s | 0.636 | 0.719 | 0.784 | 0.79 | 0.694 | 0.54 |
| 5 MB/s | 0.623 | 0.692 | 0.729 | 0.633 | 0.497 | 0.345 |
| 3 MB/s | 0.511 | 0.56 | 0.576 | 0.457 | 0.339 | 0.224 |
| 1 MB/s | 0.405 | 0.439 | 0.444 | 0.334 | 0.24 | 0.155 |

Page Size (KB)

**Figure 7**. (a) The left graph shows the utility of index pages versus page size for various index entry sizes using a high-performance disk (10ms latency, 10 MB/s transfer rate). (b) The graphs at right use a fixed-sized 16-byte entry and show the impact of disk performance on optimal page size. For high-performance disks, the optimum index page size grows from 8KB to 64KB.

These metrics become price/performance metrics when combined with the device rent (depreciated over 3 years). The Scan metric becomes a measure of the rent for a terabyte of the media while the media is being scanned. Table 8 displays these metrics for current devices:

**Table 8:** Performance Metrics of high-performance devices circa 1997.

| | **RAM** | **Disk** | **Tape robot** |
|---|---|---|---|
| **Unit capacity** | 1GB | 9 GB | 14 x 35 GB |
| **Unit price $** | 15,000$ | 2,000$ | 10,000$ |
| **$/GB** | 15,000 $/GB | 222$/GB | 20 $/GB |
| **Latency (ms)** | 0.1 micro sec | 10 milli sec | 30 sec |
| **Bandwidth** | 500 MBps | 5 MBps | 5 MBps |
| **Kaps** | 500,000 Kaps | 100 Kaps | .03 Kaps |
| **Maps** | 500 Maps | 4.8 Maps | .03 Maps |
| **Scan time** | 2 seconds | 30 minutes | 27 hours |
| **$/Kaps** | 0.3 nano $ | 0.2 micro $ | 3 milli $ |
| **$/Maps** | .3 micro $ | 4 micro $ | 3 milli $ |
| **$/TBscan** | .32 $ | 4.23$ | 296$ |

## 4. Summary

The fact that disk access speeds have increased ten-fold in the last twenty years is impressive. But it pales when compared to the hundred-fold increase in disk unit capacity and the ten-thousand-fold decrease in storage costs (Figure 1). In part, growing page sizes sixteen-fold from 512 bytes to 8 KB has ameliorated these differential changes. This growth preserved the five-minute rule for randomly accessed pages. A one- minute rule applies to pages used in two-pass sequential algorithms like sort. As technology advances, secondary storage capacities grow huge. The *Kaps*, *Maps*, and *Scans* metrics that measure access rate and price/access are becoming increasingly important.

## 5. Acknowledgments

Paul Larson, Dave Lomet, Len Seligman and Catharine Van Ingen helped us clarify our presentation of optimum index page sizes. The *Kaps*, *Maps*, and *Scans* metrics grew out of discussions with Richie Larry.